\documentclass[10pt]{iopart}
\usepackage[en-US]{datetime2}
\usepackage[english]{babel}
\usepackage{graphicx}
\usepackage{layouts}
\usepackage{siunitx}
\usepackage[hyphens]{url}
\usepackage{hyperref}
\usepackage[hyphenbreaks]{breakurl}
 \DeclareSIUnit\atomicmassunit{u}
\usepackage{nicefrac}

\begin{document}

\title[]{Integration of maXs-type microcalorimeter detectors for high-resolution x-ray spectroscopy into the experimental environment at the CRYRING@ESR electron cooler}

\author{Ph.~Pfäfflein, S.~Bernitt, Ch.~Hahn, M.~O.~Herdrich, F.~M.~Kröger, E.~B.~Menz, T.~Over, B.~Zhu, and Th.~Stöhlker}

\address{Helmholtz Institute Jena, Fröbelstieg~3, 07743~Jena, Germany}
\address{GSI Helmholtzzentrum für Schwerionenforschung, Planckstraße~1, 64291~Darmstadt, Germany}
\address{Institute for Optics and Quantum Electronics, Friedrich Schiller University, Max-Wien-Platz~1, 07743~Jena, Germany}

\author{G.~Weber}

\address{Helmholtz Institute Jena, Fröbelstieg~3, 07743~Jena, Germany}
\address{GSI Helmholtzzentrum für Schwerionenforschung, Planckstraße~1, 64291~Darmstadt, Germany}

\author{S.~Allgeier, M.~Friedrich, D.~Hengstler, P.~Kuntz, and  A.~Fleischmann}

\address{Kirchhoff Institute for Physics, Heidelberg University, Im Neuenheimer Feld~227, 69120~Heidelberg}

\author{Ch.~Enss}
\address{Kirchhoff Institute for Physics, Heidelberg University, Im Neuenheimer Feld~227, 69120~Heidelberg}
\address{Institute for Data Processing and Electronics, Karlsruhe Institute of Technology, Hermann-von-Helmholtz-Platz~1, 76344~Eggenstein-Leopoldshafen}

\author{A.~Kalinin, M.~Lestinsky, B.~Löher, and U.~Spillmann}

\address{GSI Helmholtzzentrum für Schwerionenforschung, Planckstraße~1, 64291~Darmstadt, Germany}

\ead{p.pfaefflein@hi-jena.gsi.de}
\vspace{10pt}
\begin{indented}
\DTMlangsetup{showdayofmonth=false}
\item[]\today
\end{indented}

\newpage

\begin{abstract}
We report on the first integration of novel magnetic microcalorimeter detectors (MMCs), developed within SPARC (Stored Particles Atomic Physics Research Collaboration), into the experimental environment of storage rings at GSI\footnote{GSI Helmholtzzentrum für Schwerionenforschung GmbH, Planckstraße~1, 64291~Darmstadt}, Darmstadt, namely at the electron cooler of CRYRING@ESR. Two of these detector systems were positioned at the \ang{0} and \ang{180} view ports of the cooler section to obtain high-resolution x-ray spectra originating from a stored beam of hydrogen-like uranium interacting with the cooler electrons. While previous test measurements with microcalorimeters at the accelerator facility of GSI were conducted in the mode of well-established stand-alone operation, for the present experiment we implemented several notable modifications to exploit the full potential of this type of detector for precision x-ray spectroscopy of stored heavy ions. Among these are a new readout system compatible with the multi branch system data acquisition platform of GSI, the synchronization of a quasi-continuous energy calibration with the operation cycle of the accelerator facility, as well as the first exploitation of the maXs detectors' time resolution to apply coincidence conditions for the detection of photons and charge-changed ions.
\end{abstract}

%
%
%
%
\ioptwocol

\section{Introduction}

X-ray spectroscopy of energetic highly charged ions in collisions with matter is an indispensable tool for investigations of relativistic interaction dynamics as well as for probing our understanding of atomic structures in the relativistic domain of high-$Z$ few-electron ions. In these collisions, x-ray emission by the projectile is occurring in a variety of processes, such as radiative recombination (RR) or radiative electron capture (REC)~\cite{Eichler2007}, as well as the radiative decay of excited states formed by charge changing interactions or collisional excitation~\cite{Anholt1986,Stoehlker1998,Ludziejewski2000,Gumberidze2019}. For atomic structure studies, the latter is of particular importance, as precision spectroscopy of radiative transitions in few-electron, high-$Z$ systems  provides unique access to the effects of bound-state quantum electrodynamics (QED) in the nonperturbative domain of extremely strong electromagnetic fields~\cite{Stohlker2008a,Gumberidze2005,Fritzsche2005,Durante2019}.

This type of studies has been significantly facilitated by the introduction of ion storage rings equipped with electron cooling, such as the Experimental Storage Ring (ESR) at GSI, Darmstadt~\cite{Franzke198718}. Cooling of the ion beam enables a reduction of the emittance as well as of the kinetic energy dispersion~\cite{Steck04}, resulting in a better control of the Doppler shift and a reduced broadening of the spectral lines emitted by the projectiles. In addition, the high repetition rates on the order of \SI{1}{\mega\hertz} combined with in-ring targets result in a much higher luminosity compared to single-pass setups. This enables the use of dilute gas targets~\cite{Kuehnel09} which provide single-collision conditions, thus yielding clean x-ray spectra. Finally, the deceleration capability of storage rings allows measurements at various velocities, in particular below the production threshold of the specific ion species (typically some \SI{100}{\mega\electronvolt\per\atomicmassunit} for the heaviest one- and two-electron ions)~\cite{Stoehlker1998,Stoehlker97,Scheidenberger199436}.

Exploiting these advantageous experimental conditions, previous x-ray studies at the ESR internal gas target have addressed a broad range of features of the emitted radiation, including the spectral distribution, the emission pattern and the polarization characteristics \cite[and references therein]{Stohlker2009}. Moreover, precision x-ray spectroscopy was also successfully performed at the ESR electron cooler~\cite{Gumberidze2005,Gumberidze2004}. This device lends itself as a well-defined free-electron target and offers advantages for certain types of measurements. However, the range of possible experimental configurations at the ESR cooler is severely limited by challenging geometrical and operational constraints. In light of this, the recently commissioned CRYRING@ESR provides an attractive alternative experimental facility for high-precision spectroscopy of highly-charged ions~\cite{Lestinsky2016}. This storage ring, which can be supplied with heavy ions via injection from the ESR, is optimized for low beam energies ranging from about \SI{10}{\mega\electronvolt\per\atomicmassunit} down to a few \SI{10}{\kilo\electronvolt\per\atomicmassunit}. Crucially, it also incorporates an electron cooler which provides a more flexible environment for x-ray spectroscopy setups, mainly due to the compact size when compared to the installations at the ESR.

Precision x-ray spectroscopy does furthermore benefit from recent advances in the development of small-volume low-temperature detectors (LTDs) for ionizing radiation, also referred to as microcalorimeters. These types of detectors are based on measuring a minute temperature increase, induced by the stopping of a single incident particle, in an absorber with a very small heat capacity. Compared to semiconductor detectors based on e.g. silicon or germanium, this novel type of detector provides an improvement of the intrinsic energy resolution by more than one order of magnitude. Indeed, when applied to x-ray spectroscopy in the few to tens of \si{\kilo\electronvolt} regime, various LTD technologies have demonstrated resolving powers significantly better than \num{1e-3}~\cite{Enss08,Kempf2018,Sikorsky2020,Oshima2008,Yamada2021}. In addition to the very promising spectroscopic performance, such detectors can provide stopping powers and spectral acceptance ranges comparable to standard semiconductor x-ray detectors.

These unique features make microcalorimeters a particularly promising type of detector system for the scientific program of the SPARC collaboration~\cite{Durante2019,Stohlker2014,Stoehlker2015680}, which focuses on research in the realm of atomic, quantum, and fundamental physics at GSI and FAIR\footnote{Facility for Antiproton and Ion Research in Europe GmbH, Planckstraße~1,
64291~Darmstadt,}. Within the collaboration, two microcalorimeter designs have been recently developed, namely the SIM-X detector~\cite{sim-x_tdr}, which is based on a compensated-doped silicon thermistor, and the maXs detector (\textbf{M}icro-Calorimeter \textbf{A}rrays for High Resolution \textbf{X}-ray \textbf{S}pectroscopy,~\cite{maxs_tdr}), which is a magnetic microcalorimeter (MMC). An overview of the recent progress of these activities is presented in~\cite{Kraft-Bermuth2018}. In particular, prototypes of both detector systems have been successfully deployed in recent test experiments at the gas jet target of the ESR at GSI~\cite{Hengstler2015,Kraft-Bermuth2017}. However, in these measurements the detectors were operating as stand-alone devices, i.e. without integration into the heterogeneous experimental infrastructure at GSI. In particular, their data acquisition system was not synchronized with the accelerator operation, nor was the acquired data merged with the data streams of other types of detectors and installations. These shortcomings, while not crucial in first proof-of-principle measurements, need to be overcome to exploit the full potential of this new class of detector systems.

\begin{figure*}
    \centering
    \includegraphics[width=.9\textwidth]{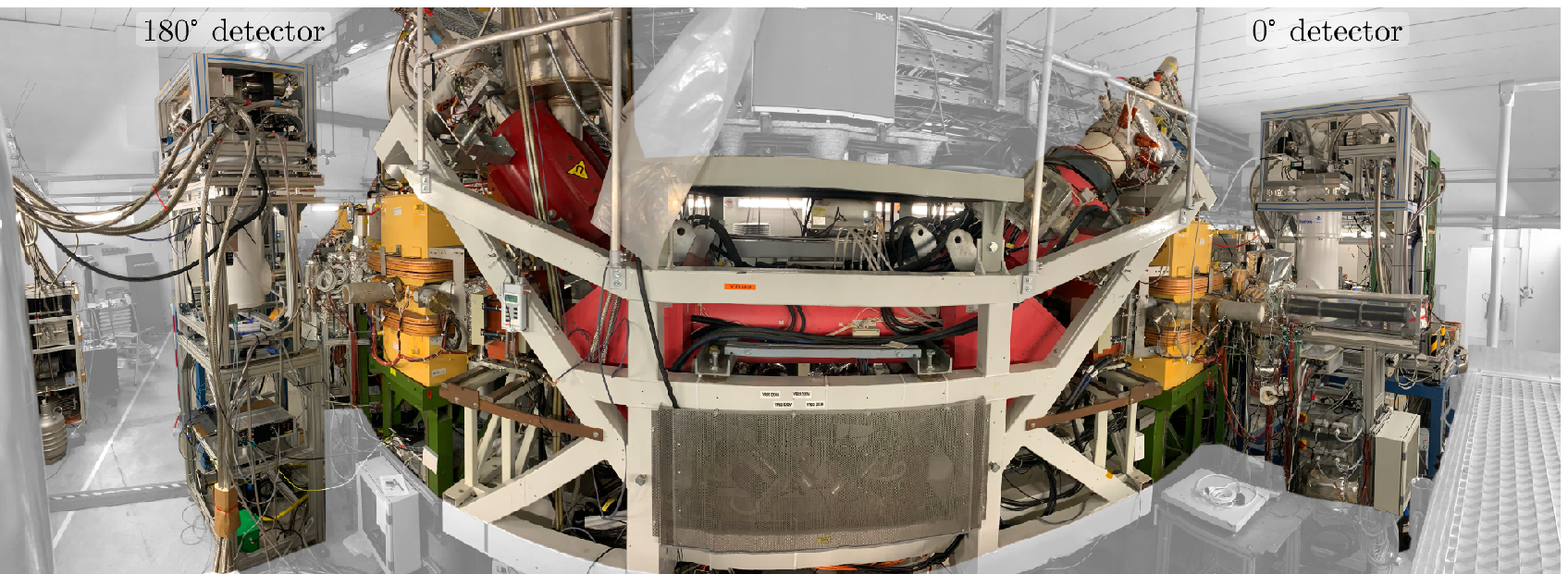}
    \includegraphics[width=.9\textwidth]{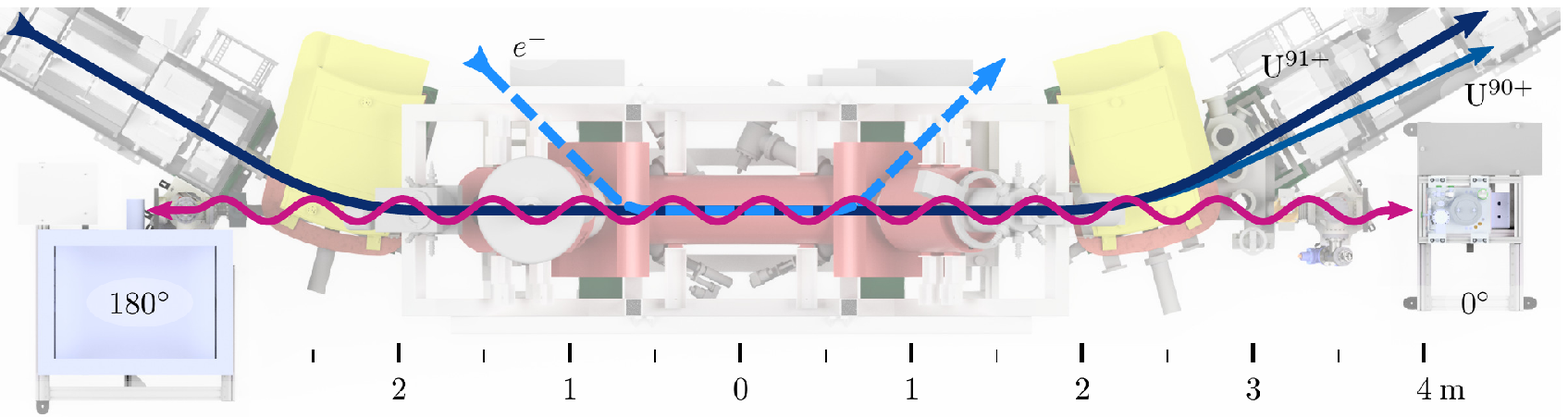}
    \caption{Top: Overview of the CRYRING@ESR electron cooler with the implemented x-ray spectroscopy setup. The electron cooler (red) in the center of the image acts as a free electron target leading to radiative recombination. X-ray photons emitted in forward and backward direction are detected by the highlighted detectors. Bottom: Schematic view of the overlapping beam paths within the accelerator. In the electron cooler the stored U$^{91+}$ beam (dark blue) is merged with an electron beam (light blue). This can lead to charge transfer in combination with the emission of a photon (violet wave). The down charged U$^{90+}$ beam is separated from the stored ions after passing through the dipole and detected by a channel electron multiplier.}
    \label{fig:overview}
\end{figure*}

In this report we present the first application of two novel maXs-type MMC detectors at the electron cooler of CRYRING@ESR, with their operation and data acquisition system being fully integrated into the experimental environment of GSI. The experiment aimed at precision spectroscopy of x-ray emission resulting from radiative recombination of stored U$^{91+}$ with cooler electrons. We will focus on the aspects that are particular to the operation of this novel type of detector system, such as the implementation of a quasi-continuous energy calibration in synchronization with the accelerator operation, as well as the first utilization of the detectors' timing capabilities to suppress background radiation via a coincidence technique.

\section{Experimental setup at CRYRING@ESR}

Recently, CRYRING has been transferred from the Manne Siegbahn Laboratory in Stockholm to the GSI/FAIR campus in Darmstadt as an in-kind contribution from Sweden to the upcoming international Facility for Anti-proton and Ion Research (FAIR) (see~\cite{Durante2019} and references therein). For this purpose the storage ring has further been optimized for future experiments with heavy bare and few-electrons ions, exotic nuclei as well as anti-protons \cite{CRYRING_tdr} that are all part of the scientific program of FAIR. The modified ring has been installed downstream of ESR as part of the modularized start version of FAIR and is now referred to as CRYRING@ESR~\cite{Geithner2017}. There it offers unique opportunities for a broad range of experiments with electron-cooled heaviest one- and few-electron ions at low energies. Since 2020, the facility has been fully commissioned both with beams from the local ion injector as well as from the ESR storage ring and is now available for first physics production runs~\cite{Zhu2022}.



\subsection{X-ray spectroscopy at the electron cooler}
The electron cooler of CRYRING@ESR features a \SI{1.2}{\meter} long cooler region where the ion and the electron beam are colinear. The cooler setup fits into a straight section of \SI{4}{\meter} between two of the ring's dipole magnets. The vacuum chambers of these magnets are equipped with view ports along the beam axis, so that detectors for photon spectroscopy can be positioned at observation angles of \ang{0} and \ang{180} with respect to the ion beam direction (see figure~\ref{fig:overview}). The detectors can be placed as close as \SI{3.5}{\meter} from the center of the electron cooler. This truly coaxial geometry enables the detection of clean spectra without distortion effects caused by delayed photon emission. This is in contrast to the situation at the ESR electron cooler, where geometrical constraints force the use of custom-made x-ray detectors mounted in retractable pockets perpendicular to the ion beam axis. These conditions lead to observation angles of $\approx$\ang{0.5} and \ang{179.5} with respect to the center of the cooler region. Delayed emission from ions passing the forward detector covers a wide range of observation angles, and the correspondingly varying Doppler shift manifests itself as low-energy tails in the recorded spectra~\cite{Reuschl2008}. Furthermore, photons below \SI{20}{\kilo\electronvolt} are effectively blocked by the used stainless steel windows which could not be circumvented in the design of the pockets. To alleviate these limitations, the view ports at the CRYRING electron cooler were recently equipped with dedicated window chambers~\cite{Kroeger2019}. These can be separated from the ring via vacuum valves and allow on-demand pumping and baking to reach ultra-high vacuum (UHV) conditions. Thus, windows optimized for the specific wavelength region of a given experiment can be mounted without breaking the ring vacuum. More specifically, for x-ray spectroscopy with photon energies down to a few \si{\kilo\electronvolt}, beryllium windows with a diameter of about \SI{30}{\milli\meter} and a thickness of \SI{100}{\micro\meter} are used.

X-ray spectroscopy studies of highly-charged ions can often significantly benefit from the use of coincidence techniques to identify those events which are associated with projectile ions undergoing charge changing processes. As the charge-modified ions are separated from the main beam in dipole magnets, this can be achieved by combining the signals of a particle detector with timing information from the x-ray detectors. A particle counter, which can be applied for this purpose, is also installed downstream of the electron cooler of CRYRING@ESR. It consists of a channel electron multiplier that amplifies secondary electrons produced by the impact of the downcharged ions on a baffle plate. This way, synchronized recording of the photon and particle detector signals allows the use of time-of-flight information, which enables a strong suppression of the background consisting of photons which are not associated with recombination events between cooler electrons and projectile ions. In addition, such time-of-flight data also offers insights into the time evolution of the emission spectrum.


The favorable experimental conditions for x-ray spectroscopy at the electron cooler of CRYRING@ESR with the described setup have recently been demonstrated using conventional high-purity germanium x-ray detectors and a stored bare lead beam~\cite{Zhu2022}. In the electron cooler section ions and cooler electrons were undergoing radiative recombination (RR), resulting in the emission of x-ray photons which carry away the binding energy plus the very small kinetic energy of the electrons relative to the ion beam. As at threshold energies a significant portion of the electrons recombine not preferentially into the ground state but into high-$n$, high-$l$ states of the projectile, the subsequent decay chains give rise to a rich and distinct pattern of characteristic x-ray transitions. Despite the short measurement time of a few days and a low beam intensity, the commissioning beam time succeeded in accumulating very clean x-ray spectra, consisting of photons from RR into Pb$^{82+}$ and subsequent characteristic transitions in Pb$^{81+}$, including the Paschen series at only a few~\si{\kilo\electronvolt} photon energy.




\section{The present experiment}

\begin{figure}
    \centering
    \includegraphics[width=.9\columnwidth]{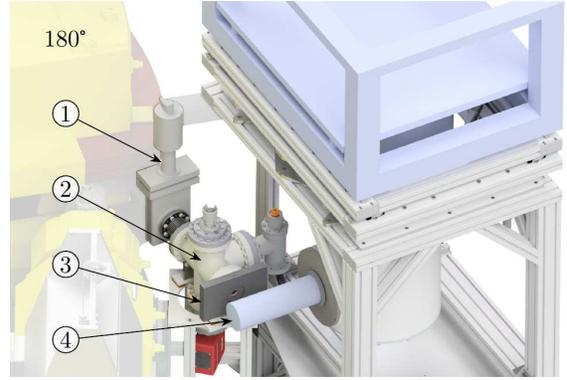}
    \caption{Detailed view of the experimental setup at \ang{180}. Highlighted are the vacuum valve (1), that couples the window chamber (2) to the beam line, as well as the calibration source holder (3) and the maXs detector system (4). Due to geometrical constraints at this position the cryostat had to be equipped with a nozzle to place the absorber array close to the view port.}
    \label{fig:detail}
\end{figure}

In an effort to push the experimental precision of spectroscopy in the range of high-$Z$ $K$-shell radiation (up to \SI{100}{\kilo\electronvolt}) below \SI{1}{\electronvolt}, we deployed the aforementioned maXs detectors for the first time at the CRYRING@ESR electron cooler. An overview of the experimental setup is presented in figure~\ref{fig:overview}. The upper photograph depicts the cooler section of the CRYRING@ESR, with maXs-type detector systems occupying the \ang{0} and \ang{180} view ports. The lower top view of the electron cooler model illustrates the trajectories of the cooler electrons and the projectile ions with their respective charge states, as well as photons emitted from the interaction region.


The following measurement cycle was applied: uranium ions were accelerated by the UNILAC/SIS18 complex to a kinetic energy of \SI{295}{\mega\electronvolt\per\atomicmassunit} and passed through a stripper foil between SIS18 and ESR to produce hydrogen-like uranium ions (U$^{91+}$). After injection into the ESR successive electron cooling and deceleration steps were applied, followed by extraction towards CRYRING@ESR at an energy of \SI{10.255}{\mega\electronvolt\per\atomicmassunit}. While up to \num{1e9} ions were injected into the ESR, this number was reduced after deceleration and transfer to CRYRING@ESR to typical values of \numrange[]{1e6}{2e6} ions. In CRYRING@ESR the beam was continuously electron-cooled at a voltage of \SI{5634.5}{\volt} and an electron current of \SI{30.5}{\milli\ampere}. The beam lifetime of \SIrange{7}{8}{\second} was determined in approximately equal proportions by charge exchange in the residual gas and RR in the electron cooler. Although the stored beam was virtually gone after about \SI{25}{\second}, the injection period was determined to approximately \SI{55}{\second} by the preparation time of the beam in the ESR. The time difference was used for calibration of the maXs detectors which is described in detail in section~\ref{sec:calibration}.


A detailed view of the spectroscopy setup located at the \ang{180} port is shown in figure~\ref{fig:detail}. The main components are the vacuum valve (1) for coupling of the window chamber (2) to the CRYRING dipole magnet, the lead-shielded box (3) that houses a variety of movable reference gamma sources for energy calibration, and finally the maXs detector system (4). Due to geometrical constraints at this view port the cryostat of the detector had to be equipped with an extension to place the absorber array as close as possible to the view port, whereas for the \ang{0} port the detector head is located inside the cryostat barrel.


\subsection{The maXs-100 MMC detector system}

\begin{figure}
    \centering
    \includegraphics[width=.9\columnwidth]{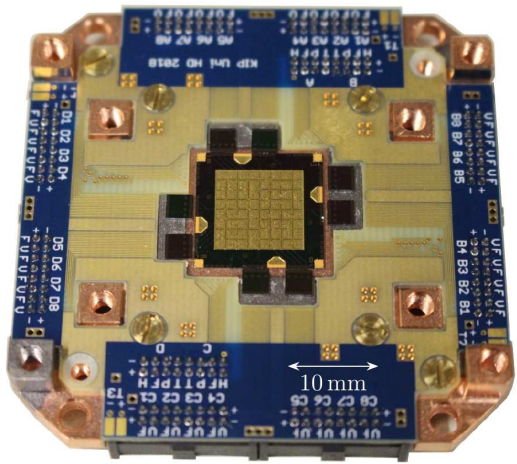}
    \caption{Photograph of one of the detector heads used in the present experiment. In the center the \numproduct{8 x 8} pixels of the maXs-100 chip are visible. They are surrounded by 8 chips containing the 32 first-stage readout SQUIDs. A collimator made of copper (not shown) is added on top of the module to shield the SQUID chips from direct exposure to the incident radiation.}
    \label{fig:maxs}
\end{figure}

For the present experiment we employed two novel maXs-100 detectors, which were developed in collaboration between Heidelberg University and Helmholtz Institute Jena~\cite{Hengstler2015,Pies2012,Herdrich2020,Weber2020}. The underlying MMC technology combines a high energy resolution with a fast intrinsic rise time and great flexibility in the choice of absorber materials~\cite{Kempf2018}. These unique characteristics make maXs-type detectors a particularly promising tool for a multitude of precision x-ray spectroscopy experiments, especially for high photon energies and broad spectral bandwidths that are challenging for crystal spectrometers.

The maXs-100 detector chip together with the first stage of readout electronics is integrated into a detector head, see figure~\ref{fig:maxs}. The modular design of the maXs detector system allows the exchange of the detector head with alternative detector arrays, being optimized for specific experimental requirements with respect to active area, stopping power and spectral resolution. The maXs-100 detector design, in particular, is optimized for spectroscopy of $K$-shell transitions (close to \qty{100}{\kilo\electronvolt}) in heavy ions, and features absorbers made of electro-deposited gold with a thickness of \SI{50}{\micro\meter}. They provide an acceptable quantum efficiency of about \qty{40}{\percent} for stopping of \SI{100}{\kilo\electronvolt} x-rays. The lateral dimensions of the absorber pixels are approximately \qtyproduct[product-units = single]{1.25 x 1.25}{\milli\meter\squared}. Thus, the complete detector array of \numproduct{8x8} pixels has an effective active area of about \qty{1}{\centi\meter\squared}.

The measurement principle is as follows: Each absorber is thermally coupled to a paramagnetic temperature sensor made of sputtered Ag:Er. A superconducting meander-shaped pickup coil made of sputter-deposited niobium is used to generate a magnetic field in the sensor volume. Upon absorption of an energetic photon, the pair of absorber and sensor is heating up and this temperature increase of the sensor results in a decrease of its magnetization. Thus, the energy deposition in the absorber is translated into a change in magnetization of the sensor material. This change is then measured by a low noise, high bandwidth SQUID magnetometer which is connected to the meander coil via a superconducting circuit.

Once the sensor heats up as a result of energy deposition $\delta E$ within the absorber, the change in magnetization is described by $A \propto \frac{1}{C} \frac{\mathrm{d}M}{\mathrm{d}T} \delta E$. Thus, the obtained signal is inversely proportional to the heat capacity $C$ of the absorber--sensor assembly, and directly proportional to the steepness of the magnetization curve $\frac{\mathrm{d}M}{\mathrm{d}T}$ of the paramagnetic sensor. To achieve a large $\frac{\mathrm{d}M}{\mathrm{d}T}$ value, the maXs detectors are operated at typical temperatures of \SI{20}{\milli\kelvin}, using $^3$He/$^4$He dilution cryostats delivered by Bluefors. To further optimize the obtainable spectral resolution, a small detector volume, with a correspondingly small $C$, proves beneficial. Larger absorbers, on the other hand, may provide a superior quantum efficiency and/or enable a larger solid angle coverage. The selected balance of this tradeoff distinguishes the various detectors of the maXs family~\cite{maxs_tdr}. For the maXs-100 detector, the desired design value of the spectral resolution of well below \SI{50}{\electronvolt} FWHM at \SI{100}{\kilo\electronvolt} incident photon energy motivated the absorber dimensions listed above.

The temporal evolution of the change in magnetization is described by an exponential rise $A(t) \propto 1-e^{-t/\tau_{\textrm{r}}}$, with the time constant $\tau_{\textrm{r}}$ being limited by the throughput of an artificial thermal link between absorber and sensor. This is in particular crucial for (comparably) large-area absorbers, such as the one of the maXs-100 with \qty{1.56}{\milli\meter\squared}, as previous tests with maXs-type detectors showed signal shapes which depended on the exact position of the x-ray absorption due to the finite thermal conductance inside the absorbers~\cite{pies2012maxs}. This may lead to a decrease in energy resolution as well as deterioration of the achievable timing information. To mitigate these effects each absorber is connected to its temperature sensor only via a weak thermal link, ensuring a complete thermalization of the deposited energy in the absorber volume before a significant amount of heat reaches the sensor. Therefore, although the MMC technology can provide very fast signal rise times down to a few \qty{100}{\nano\second}, the design value for the maXs-100 detector was set to a significantly longer time constant of \qty{8}{\micro\second}.

\begin{figure}
    \centering
    \includegraphics[width=.9\columnwidth]{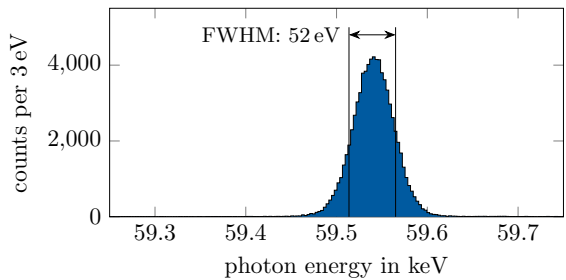}
    \caption{The \qty{59.54}{\kilo\electronvolt} gamma line of $^{241}$Am as recorded by one of the two maXs-100 detectors. The spectrum results from adding up the data obtained from the individual pixels. Note that this data was measured in a dedicated laboratory room to characterize the detector performance in the absence of disturbances caused by the accelerator environment, before the actual experiment at CRYRING@ESR was performed.}
    \label{fig:spectrum}
\end{figure}

To reduce the sensitivity to external $B$ field fluctuations and to minimize cross-talk in the close-packed array, two neighboring detector pixels are gradiometrically coupled and read out by a single SQUID channel. A decrease of the magnetization of one pixel results in an increase of the output signal level, while a decrease of the magnetization of the other pixel leads to a decrease of the output signal. Therefore identical changes of the magnetization -- caused e.g. by a change in the base temperature -- of two coupled ideal pixels compensate each other and maintain a stable baseline. Energy deposition in one of the absorbers, however, creates a temperature difference between the pixels and results in a significant change of the output voltage. For the readout of the maXs detectors, the signals from the first SQUID stage are amplified by a second SQUID stage and then fed to a XXF-1 SQUID electronics from MAGNICON\footnote{Magnicon GmbH, Barkhausenweg 11, 22339 Hamburg}, which are located outside of the cryostat. Through a flux-locked loop feedback system, the magnetic flux inside the first-stage SQUID is kept constant, resulting in a voltage signal that changes linearly with the (net) magnetization of the sensors connected to the specific readout channel. Finally, the output signal of the SQUID electronics is digitized for further analysis, see \ref{sec:daq} for details.


Even though the baseline level is stabilized, the pulse amplitude upon absorption of a photon is still affected by the base temperature as both, the heat capacity and the magnetization curve of the sensor, are temperature-dependent quantities. Thus, to achieve a satisfactory energy resolution a correction of the amplitude with respect to the varying chip temperature is vital. Therefore, the pixel pairs of four of the 32 detector channels are coupled to the readout SQUID in an asymmetric way, thus avoiding the aforementioned cancellation. As a consequence, the resulting baseline voltage of these four channels is proportional to the chip temperature and can be used for monitoring its time evolution. By monitoring the voltage level of these channels for each recorded x-ray event, fluctuations of the temperature can be quantified and corrected for in a reliable way. However, to track slow drifts of the total gain and to verify the gain stability, frequent or even continuous reference measurements with calibration sources are crucial.

\begin{figure}
    \centering
    \includegraphics[width=.9\columnwidth]{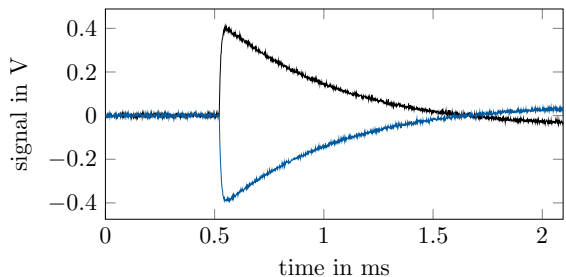}
    \caption{Two typical pulses from one of the maXs-100 detector channels, resulting from the absorption of \SI{59.54}{\kilo\electronvolt} photons from $^{241}$Am. The signal from one of the two sensors connected to the readout channel goes in the positive direction, while the other sensor produces a signal in the opposite direction. This results in a cancellation of baseline drifts when the detector base temperature is changing. See text for details.}
    \label{fig:pulse}
\end{figure}

After their arrival at GSI, the spectral performance of the detectors was evaluated in a laboratory on-site, using various well-characterized gamma sources. Storing the complete pulse trace for each event is key to achieving the best possible spectral resolution, because it permits the application of pulse shape algorithms tailored to the characteristics of each individual pixel. After processing the signals from each pixel separately, the spectra were co-added and the overall resolution was determined. As shown in figure~\ref{fig:spectrum}, the \SI{59.54}{\kilo\electronvolt} gamma line of $^{241}$Am was observed at a resolution of slightly above \SI{50}{\electronvolt} FWHM. While this is a satisfactory value, we noticed that the sensitive read-out of the detectors did pick up interferences caused by the nearby accelerator complex, in particular the operation of the fast-ramping kicker magnets. This observation, together with the fact that the cooling systems of the cryostats are prone to temperature drifts as a result of vibrations, led to the expectation of an inferior spectral resolution in the actual experimental environment. There, the detectors were to be mounted at a height of \qty{2}{\meter} on 3D moveable support frames~\cite{Pfaefflein2020} in close vicinity to the accelerator structures. Preliminary data analysis suggests that both detectors achieved energy resolutions of \qtyrange{80}{90}{\electronvolt} FWHM or better, when operating at CRYRING@ESR. Detailed data analysis is still ongoing.


\subsection{Data acquisition system}
\label{sec:daq}
The output signal of the SQUID electronics was digitized at a sampling rate of \qty{125}{\mega\hertz} with 16-bit resolution over an input range of $\pm$ \SI{2.5}{\volt}. This was realized by using 16-channel SIS3316 modules from STRUCK Innovative Systems. Figure~\ref{fig:pulse} shows two typical pulses from one of the detector channels, resulting from the absorption of \SI{59.54}{\kilo\electronvolt} photons from $^{241}$Am. As mentioned above, the signal from one of the two sensors connected to this channels leads to a positive signal, while the other has an inverse polarity. The signal rise is determined by the thermalization time of the absorber-sensor pair, which for the maXs-100 detector is on the order of \qty{10}{\micro\second}. The decay constant of the signal is determined by the time necessary for the absorber and sensor to reach thermal equilibrium with the thermal bath, which happens on the \unit{\milli\second} time scale. As a consequence, a reduced sampling period of about \qty{100}{\nano\second} is still sufficient to capture all relevant features of the detector signal. This was achieved by averaging over 16 samples which reduces the amount of stored data, while at the same time acting as a low pass filter suppressing high frequency noise compared to a simple reduction of the sampling rate. This way, for every readout channel a trace of $2^{14}$ 16-bit data points was written to the internal buffer of the digitizer modules, covering a time window of approximately \qty{2}{\milli\second} at an effective sampling rate of \qtyproduct[product-units = single]{16 x 8}{\nano\second}.


\begin{figure}
    \centering
    \includegraphics[width=.9\columnwidth]{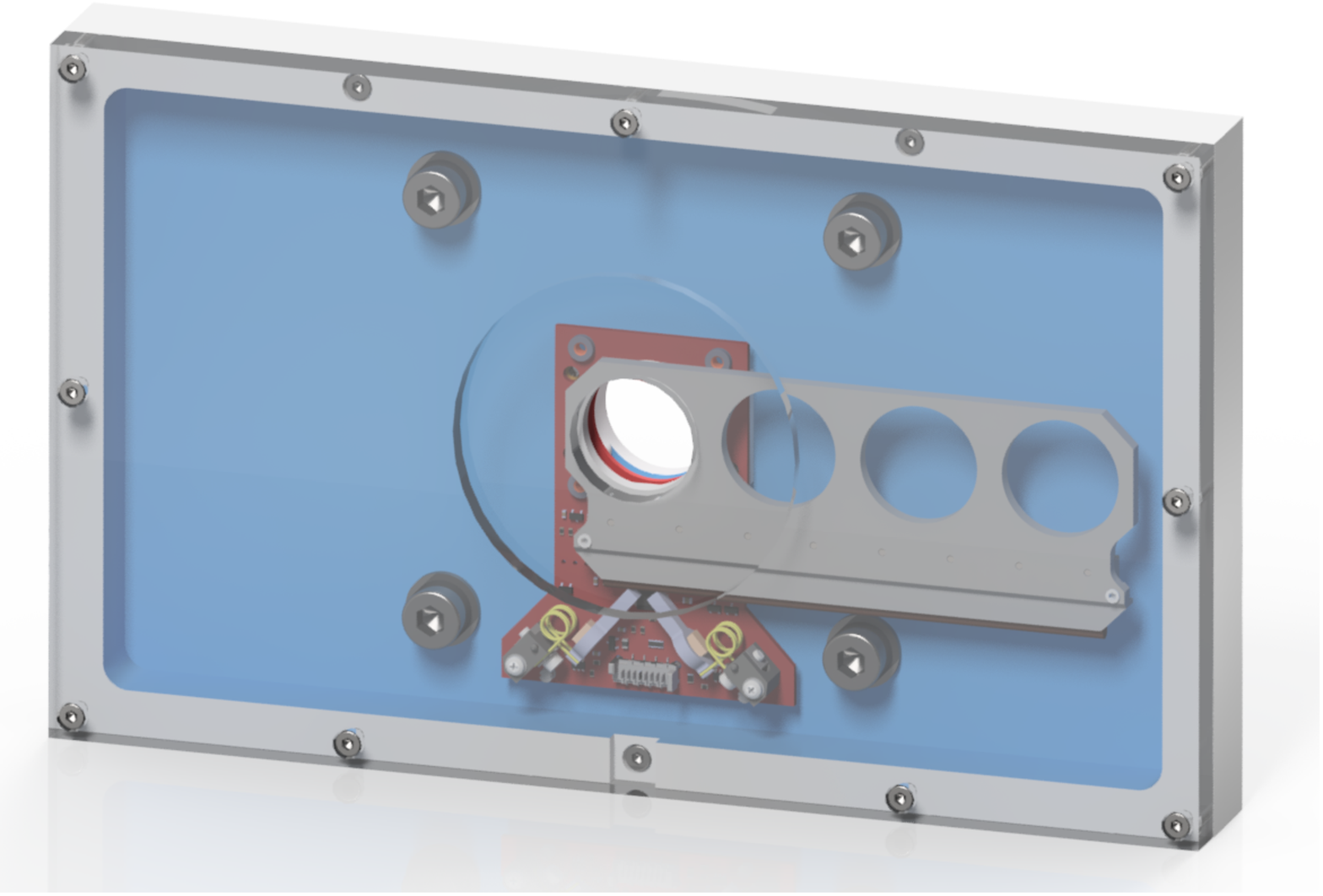}
    \vskip0.5cm
    \includegraphics[width=.9\columnwidth]{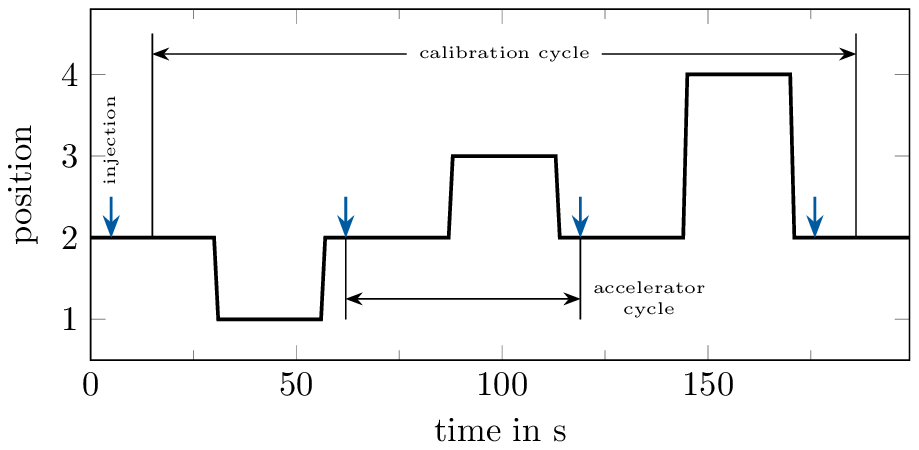}
    \caption{Upper part: Calibration source holder shown in its enclosure. Three of the four mounting positions were equipped with gamma sources, which were placed in front of the detector in synchronisation with the accelerator operation as shown in the lower part. When the beam was injected the open position was used for \qty{25}{\second} to record radiation emitted from the CRYRING electron cooler. After the beam intensity had dropped by about one order of magnitude, one of the gamma reference sources was placed in front of the detector.}
    \label{fig:calibration}
\end{figure}

\begin{figure*}[htb]
    \centering
    \includegraphics[width=\textwidth]{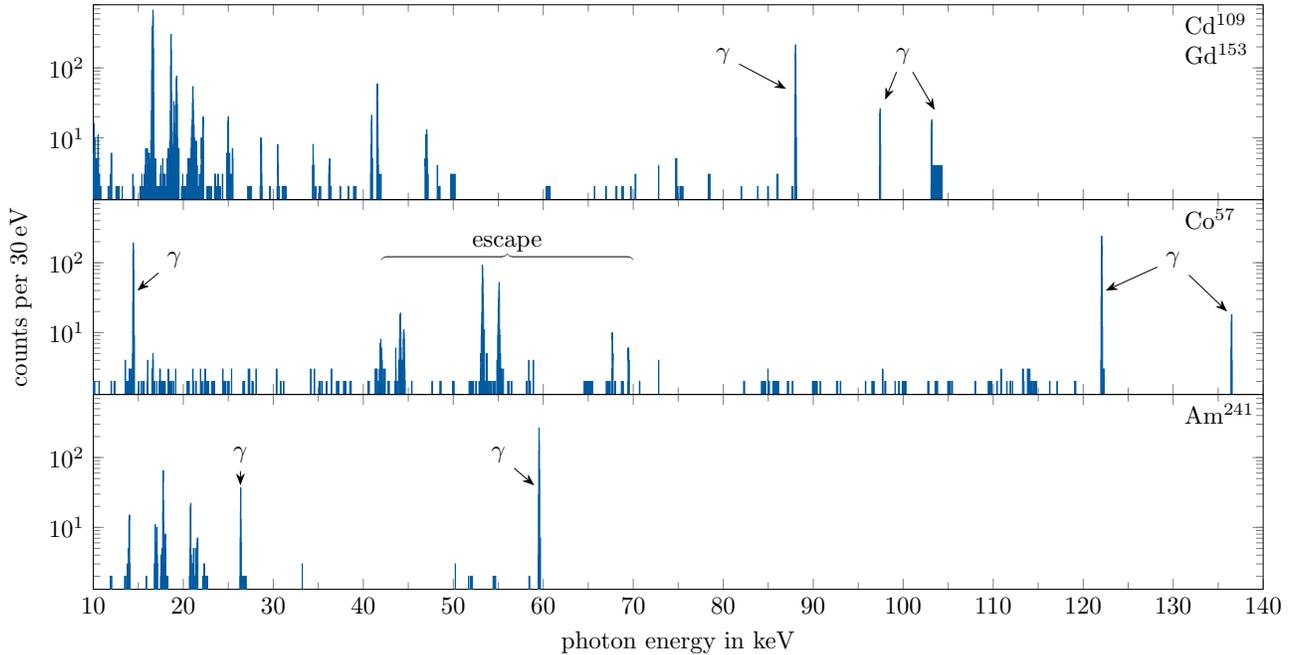}
    \caption{Calibration spectra recorded by a typical pixel of the \ang{180} detector over a period of \qty{24}{\hour} of uninterrupted accelerator operation. For the main gamma lines of interest, namely \qty{59.5}{\kilo\electronvolt} from $^{241}$Am, \qty{88}{\kilo\electronvolt} from $^{109}$Cd and \qty{122}{\kilo\electronvolt} from $^{57}$Co, more than 600 counts were accumulated. Assuming a resolution of \qty{50}{\electronvolt} FWHM, this results in an expected statistical uncertainty of the centroid determination of less than \qty{1}{\electronvolt}. The other features in the spectrum are mainly due to transitions in the atomic shells of the sources' material as well as to escape events, if the incident photon energy is above the $K$ threshold of the gold absorber. See text for details.}
    \label{fig:spectracalibration}
\end{figure*}

For consistency checks, each SQUID signal was split and digitized by two different Struck ADCs, which were part of independent data acquisition systems (DAQ). One of the two systems was based on an established stand-alone scheme~\cite{Mantegazzini2021}. The second ADC was connected to a standard GSI DAQ system~\cite{Kroeger2020} running the multi-branch system (mbs) software architecture, which is described in the following. Whenever a pulse from the maXs detector was registered by the built-in trigger logic of the SIS3316 modules, the traces of all channels that exhibited a significant deviation from the baseline level were read out by the DAQ. The data was written to file on an event by event basis, using the list-mode data format. To this end, the digitizers were integrated into a system based on NIM\footnote{Nuclear Instrumentation Standard}- and VME\footnote{Versa Module Eurocard}-type electronic modules. This allows information from the heterogeneous experimental environment of GSI/FAIR to be easily merged with the photon detector data, using well-established hardware and software solutions. In the present case, the data stream from the maXs detectors was augmented with information on the accelerator operation, such as the trigger signal indicating a new injection into CRYRING@ESR as well as a GSI-wide time stamp from the General Machine Timing System. Moreover, with each photon event also the signal from the particle detector downstream from the electron cooler section was recorded over a time window of \qty{0.5}{\milli\second}. For the fast particle signal a faster effective sampling rate of \qtyproduct[product-units = single]{4 x 8}{\nano\second} was chosen.

\subsection{Quasi-continuous energy calibration}\label{sec:calibration}

Apart from the temperature increase upon absorption of an incident photon, the signal characteristics of the maXs detectors are also determined by subtle variations of the base temperature of the detector chip as well as by specific parameters of the readout SQUIDs, such as their working points, which can be subject to instabilities. Exploiting the full potential of the detector thus relies on the ability to identify and correct for these effects. In particular, to trace variations in the detector response and thus achieve the optimum spectral precision, it is necessary to record photons from well-defined reference gamma sources throughout the whole measurement. However, since the heat load on the detector chip initiated by a high-flux gamma source can already be sufficient to change the detector temperature, it is not possible to interleave the measurement schedule with a series of short time windows during which spectra from reference sources are taken at high impact rates. For the specific type of detectors in the present experiment, the reference source activities were chosen such that the incident rate per detector channel was limited to \SI{1}{\hertz} to avoid a significant alteration of the detector response by the incident photons. At the same time, the statistical uncertainty for the line positions of the various reference lines for each calibration run was desired to be better than \SI{1}{\electronvolt}. Considering these aspects together with the aforementioned detector resolution, a significant fraction of the total measurement time had to be dedicated to accumulating photons from the reference sources. Still, obtaining the cleanest possible spectra from the projectile ions precluded the use of permanently installed reference sources, whose spectra would overlap with the relevant spectral features of the radiation being emitted in the interaction of the stored ions with the cooler electrons.

\begin{figure*}[htb]
    \centering
    \includegraphics[height=0.45\textwidth]{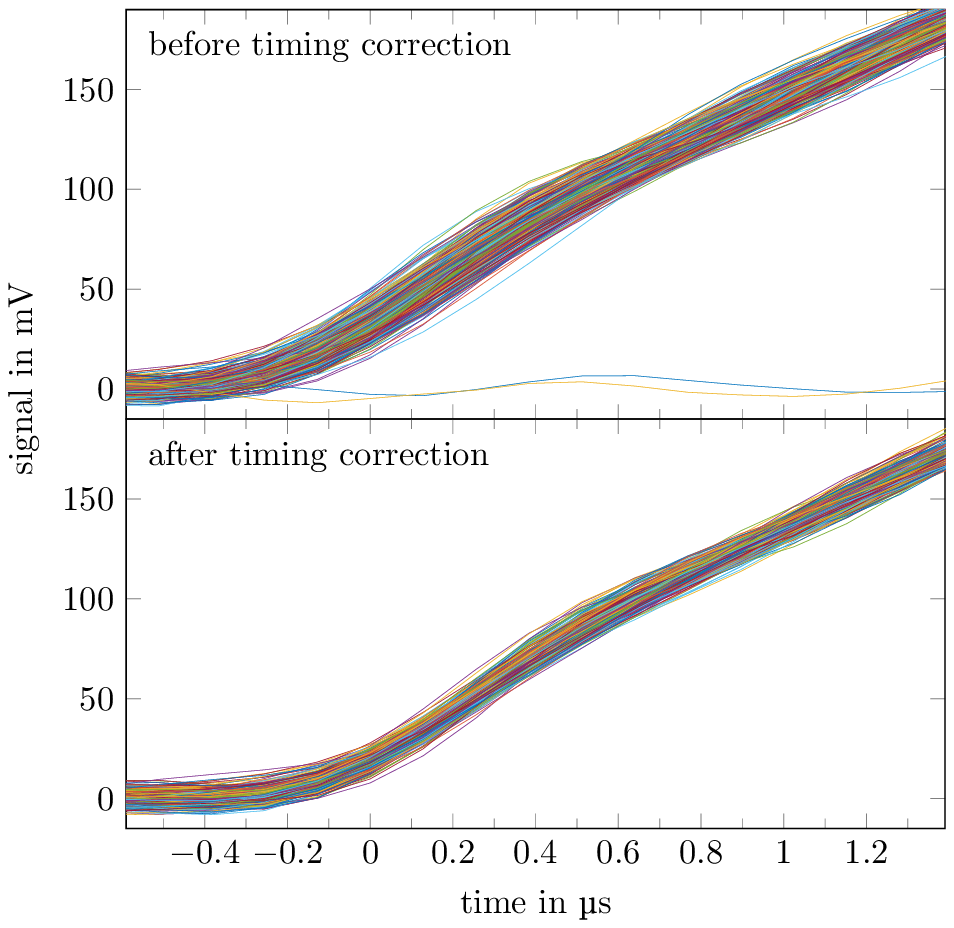}
    \hspace{.5cm}
    \includegraphics[height=0.45\textwidth]{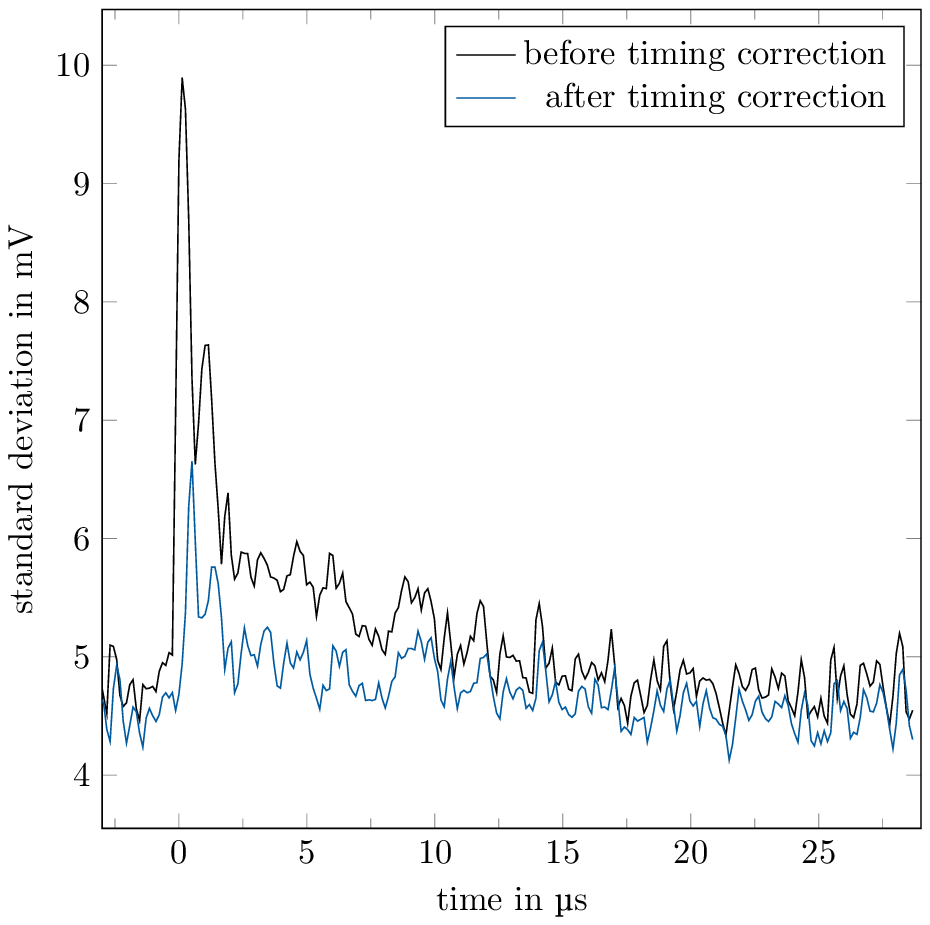}
    \caption{Timing behavior of \num{300}~typical detector pulses from one of the \num{64}~absorber pixels, resulting from \qty{122}{\kilo\electronvolt} $^{57}$Co gamma photons impinging on the detector at \ang{180}. Left: Pulses before (upper part) and after (lower part) the application of an algorithm that shifts the moment of takeoff to the zero point on the time axis. Right: In the point-wise standard deviation a timing mismatch appears as a peak in the region after the takeoff point of the pulses where the slope is the steepest. The homogenization procedure results in a strong reduction of this feature.}
    \label{fig:timecomp}
\end{figure*}

To achieve a quasi-continuous exposure of the maXs detectors to reference radiation while at the same time avoiding a mixing of these photons with the emission from the projectile ions, the sources were mounted on a motorized holder (ELL9 from Thorlabs) whose movement was synchronized with the accelerator operation. A CAD model of the source holder inside its lead-shielded housing is shown in the upper part of figure~\ref{fig:calibration}. The complete assembly was mounted in front of the view ports of the window chambers described above, see also figure \ref{fig:detail}. The holder provides four frames, of which three were equipped with $^{241}$Am, $^{57}$Co and $^{153}$Gd sources, respectively. These were chosen because of their well-defined, intense gamma lines whose energies are known to a precision of \SI{1}{\electronvolt} or better, covering the region from roughly \SI{14}{\kilo\electronvolt} to \SI{136}{\kilo\electronvolt}. This range coincides with the most intense features of the projectile radiation, namely the $K$-RR, $L$-RR, and $M$-RR radiation as well as the series of $L$, $M$, \textellipsis $\rightarrow$ $K$ and the $M$, $N$ \textellipsis $\rightarrow$ $L$ transitions. At \ang{180} the $^{153}$Gd source was accompanied by an additional $^{109}$Cd source, whose \SI{88}{\kilo\electronvolt} line is close to the red-shifted $K_{\alpha}$ radiation of U$^{90+}$. Each of the sources was equipped with an absorber foil of a material and thickness being optimized to suppress the irrelevant atomic transitions, thus maximizing the fraction of the gamma lines of interest within the source spectrum. Finally, one of the source holder's frames, labeled as number~2, was left empty so that photons from the ion--electron interaction inside the cooler could reach the detector unaffected.

The movement of the holder was synchronized with the injection trigger of the accelerator and the operation pattern, as illustrated in the lower part of figure~\ref{fig:calibration}, was as follows: After each beam injection, the empty frame (number 2) was in front of the detector for \SI{25}{\second}. At the end of this period, with the beam intensity already reduced by more than an order of magnitude, the gamma sources mounted in the frames with numbers 1, 3, and 4, respectively, were successively placed in front of the detector for another \SI{25}{\second}. Finally, the holder was moved back to the empty frame position to await the next injection into CRYRING@ESR. For each photon event recorded by the maXs detectors, the corresponding position of the gamma source holder was read out by the data acquisition system. This way almost \qty{50}{\percent} of the total measurement time could be devoted to taking data with the reference sources. 

The spectra recorded by a typical pixel of the maXs detector at \ang{180} are presented in figure~\ref{fig:spectracalibration}. This data was accumulated over a period of \qty{24}{\hour} of uninterrupted accelerator operation and the photon events were filtered according to the frame position at the time of recording. Aside from the (gamma) lines of interest, the spectra consist mainly of radiative transitions in the atomic shells of the sources' materials, as well as of so-called escape events: the small pixels of microcalorimeters are especially prone to $K$-shell transitions escaping from the absorber volume, resulting in an incomplete detection of the incident photon energy. In the present case we found an escape fraction of approximately \qty{60}{\percent} for incident photons with energies above the $K$ threshold of the gold absorbers, which is about \qty{80.7}{\kilo\electronvolt}. This finding is in agreement with simulations based on the EGS5 package that was already successfully applied to model the efficiency of semiconductor detectors~\cite{Weber2011}. Combining the fraction of escape events with the photoabsorption cross section of gold, this results in a photopeak efficiency of about \qty{15}{\percent} at \qty{100}{\kilo\electronvolt} incident photon energy. Currently a \SI{100}{\micro\meter} thick version of the maXs-100 absorber is in development, which is expected to yield an improved efficiency of close to \qty{35}{\percent} for this energy.

The presence of these escape lines, together with the contribution of atomic transitions that could not be completely suppressed by the aforementioned filter foils, limited the fraction of usable gamma photons in the calibration spectra to values between \qty{10}{\percent} and \qty{50}{\percent}. Nevertheless, for each of the main lines of interest, namely \qty{59.5}{\kilo\electronvolt} from $^{241}$Am, \qty{88}{\kilo\electronvolt} from $^{109}$Cd and \qty{122}{\kilo\electronvolt} from $^{57}$Co, more than 600~counts are found in the photopeaks. For the design value for the spectral resolution (better than \qty{50}{\electronvolt} FWHM) this results in expected statistical uncertainties of the peak centroid determinations of less than \qty{1}{\electronvolt}. When making a worst-case assumption of only \qty{100}{\electronvolt} FWHM detector resolution, the corresponding uncertainty of the reference peaks would still be below \qty{2}{\electronvolt}. Thus, for each day of the measurement an individual high-quality energy calibration can be provided for every pixel.

\section{Time resolution for coincidence measurements}
As described above, x-ray spectroscopy studies can often significantly benefit from the use of coincidence techniques to discriminate the photons of interest from unrelated radiative processes. In ion storage rings, when the photon emission is accompanied by a change in the projectile charge, this can be achieved by combining the signals of a particle detector with timing information from the photon detectors. However, to the best of our knowledge the extraction of time information from the pulses of MMC detectors was never demonstrated before. In this section we describe the necessary steps to obtain time information from the maXs detectors and relate this information to the hit time of the particle counter. This is finally applied to set a coincidence condition on the recorded spectral data.


\begin{figure}
    \centering
    \includegraphics[width=.9\columnwidth]{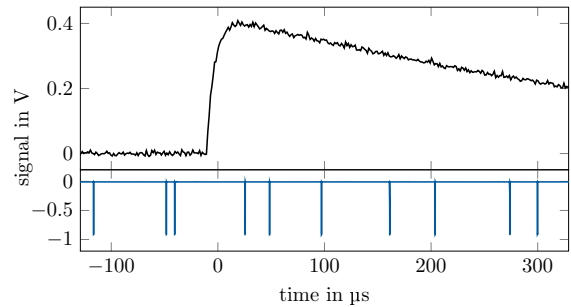}
    \caption{Trace of a photon pulse from the maXs-100 detector shown together with a logic signal indicating hits on the particle detector. See text for details.}
    \label{fig:time_of_flight}
\end{figure}

Figure~\ref{fig:timecomp} illustrates the timing behaviour of \num{300}~typical detector pulses caused by the absorption of \qty{122}{\kilo\electronvolt} $^{57}$Co gamma photons, collected from one of the \num{64}~absorber pixels of the detector at \ang{180}. On the left side the pulses are plotted before and after the application of an algorithm to homogenize the starting time of the pulses, respectively. Note that purely for illustrative purposes a moving-average filter was applied to the pulses, which suppresses high-frequency fluctuations. In the upper part the time axis for each pulse is defined according to the trigger time of the built-in finite impulse response (FIR) filter of the digitizer modules which also initiates the readout of the data acquisition system. This filter was designed for significantly faster pulse rise times, and for the present data results in a timing jitter of up to several \qty{100}{\nano\second}, i.e. not all pulses leave the baseline at the zero point. Moreover, the presented data include two pulses which take off much later than the time window depicted in the figure. These were recorded within a readout cycle which was triggered by a previous pulse on another channel, and had its time axis therefore already defined.

In the lower left part of figure~\ref{fig:timecomp} the same pulses are presented after a correction of their positions on the time axis was performed. This was done by first applying an algorithm that mimics the functionality of a constant fraction discriminator (CFD) and, in a second step, by matching the time offsets of the individual pulses to a template pulse that was obtained by averaging over a large number of individual pulses. The right side of figure~\ref{fig:timecomp} shows the effect of this homogenization on the point-wise standard deviation of the individual signal levels. When the pulses are not starting at the same point in time, the signal levels strongly deviate from each other in the region after the takeoff point of the pulses, where their slope is steepest. This feature, clearly visible for the uncorrected data set, is significantly reduced by the homogenization algorithm. After its application, the arrival time of the photon signal is therefore well defined.

\begin{figure}
    \centering
    \includegraphics[width=.9\columnwidth]{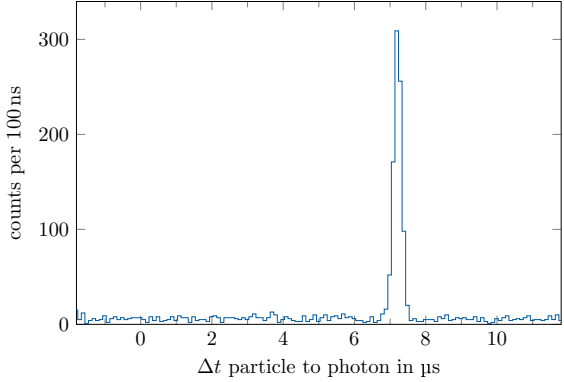}
    \caption{Difference in the arrival time of photons at the \ang{180} maXs detector and corresponding hits on the particle detector. The peak is due to the fact that photons which are emitted as a result of projectile ions undergoing radiative recombination with the cooler electrons exhibit a fixed time relation to the detection of the respective ions by the particle detector. See text for details.}
    \label{fig:tof_spec}
\end{figure}

Due to technical constraints the pulses of the particle detector could not be directly fed into the digitizer modules. Instead, its analog output was first converted into a negative NIM-type logical signal, using a conventional CFD module, which was then fed into a separate SIS3316 module. For this module a faster effective sampling rate of \qtyproduct[product-units = single]{4 x 8}{\nano\second} was applied as the logical signal had a length of less than \qty{200}{\nano\second}. To obtain the time of flight data, for each photon event the arrival time of the maXs pulse was then related to each of the particle detector hits as shown in figure~\ref{fig:time_of_flight}.

A histogram of the differences in arrival times between the photons and the corresponding particle detector signal is presented in figure~\ref{fig:tof_spec}. For the data not being dominated by random background photons, in the figure only those events with a photon energy in close proximity to the expected line positions of the projectile radiation are displayed. The clear coincidence peak with a FWHM of less than \qty{400}{\nano\second} indicates that with the present experimental setup it is indeed possible to exploit the time resolution of maXs-type detectors for applying a coincidence condition on the observed photons. However, the peak position of close to \qty{7}{\micro\second} is only partially explained by the projectile time of flight from the point of photon emission, i.e. the interaction zone with cooler electrons, to the position of the particle detector (a few \qty{100}{\nano\second}). Instead the observed time difference is mainly caused by a delayed triggering of the digitizer module recording the particle detector signal. Due to the high rate of the particle detector of up to \qty{100}{\kilo\hertz} for the most intense beam injections, the readout was only started once a photon pulse in one of the maXs detector channels was detected, which were digitized in two separate modules. Thus, the time axis of the recorded particle signal is defined in relation to the triggering of the FIR filter of the respective module registering the photon pulse. The relatively slow rise time of the maXs pulses resulted in a time gap between the actual time of arrival of the photon pulse and the release of the trigger signal by the built-in FIR filter of a few \unit{\micro\second}. However, for selecting only those photons which originate predominately from projectile ions undergoing radiative recombination with the cooler electrons, it is sufficient to set a condition on the observed coincidence peak.

\begin{figure}
    \centering
    \includegraphics[width=.9\columnwidth]{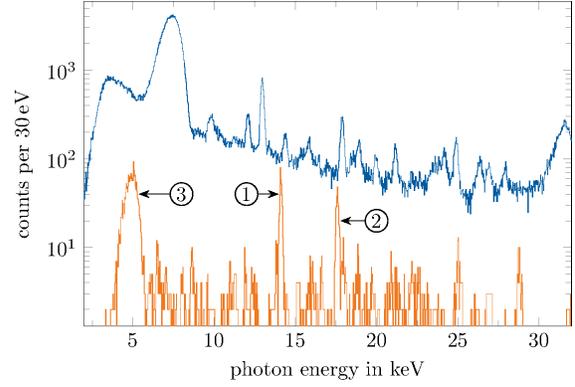}
    \caption{X-ray spectra recorded by the \ang{180} maXs detector at the electron cooler of CRYRING@ESR demonstrating the feasibility of coincidence measurements with MMC-based microcalorimeters. The spectrum containing all photons while no calibration source was in front of the detector (blue) is compared to the coincident spectrum (orange). The marked peaks correspond to the $[1s_{1/2},3d_{5/2}]\rightarrow[1s_{1/2},2p_{3/2}]$ (1) and $[1s_{1/2},3d_{3/2}]\rightarrow[1s_{1/2},2p_{1/2}]$ (2) transitions, respectively; (3) arises from bremsstrahlung. See text for details.}
    \label{fig:coinc_spec}
\end{figure}

The result of the coincidence technique for background suppression is illustrated in figure~\ref{fig:coinc_spec}. There we present x-ray spectra recorded under \ang{180}, with and without requiring the photons to be accompanied by the detection of a particle detector hit within the coincidence peak. The spectrum without the coincidence condition consists of those events which were recorded without having a reference source in front of the detector, and is at every point dominated by a broad background with distinct peaks, mainly due to fluorescence radiation from the various materials in the experimental setup. In the coincident spectrum, on the other hand, various peaks at the expected line positions of the faint projectile radiation are visible. 



\section{Conclusion}
In this work we reported on the first integration of low-temperature, small-volume detectors for precision x-ray spectroscopy in an experimental environment that is typical for heavy ion studies at storage rings, namely the operation of two maXs-type detectors at CRYRING@ESR at GSI/FAIR. This novel type of detector features a unique combination of a high spectral resolution, comparable to crystal spectrometers, with the broad bandwidth acceptance of solid-state detectors. In contrast to previous test measurements on the GSI campus, where the microcalorimeters were operating in a stand-alone mode, the present experiment implemented several notable modifications to exploit their full potential for precision x-ray spectroscopy of stored heavy ions. Among these are a new readout system compatible with the multi branch system data acquisition platform of GSI, the synchronization of the quasi-continuous energy calibration with the operation cycle of the accelerator facility as well as the first exploitation of the maXs detectors' time resolution. The successful demonstration of these adaptions paves the way for a broad range of precision x-ray spectroscopy studies.

\section{Acknowledgment}
The authors are indebted to the local teams at GSI, in particular of ESR and CRYRING@ESR, for providing us with an excellent beam. This research has been conducted in the framework of the SPARC collaboration, experiment E138 of FAIR Phase-0 supported by GSI. It is further supported the European Research Council (ERC) under the European Union’s Horizon 2020 research as well as by the innovation program (Grant No. 824109 “EMP”). B. Zhu acknowledges CSC Doctoral Fellowship 2018.9 - 2022.2 (Grant No. 201806180051). We also acknowledge the support provided by ErUM FSP T05 - “Aufbau von APPA bei FAIR” (BMBF n° 05P19SJFAA and n° 05P19VHFA1).

\section{References}
\bibliography{references}{}
\bibliographystyle{iopart-num}
\end{document}